\documentclass{PoS}

\newcommand{\ex}{{\rm exp}}
\newcommand{\uz}{[ U_0 ]}
\newcommand{\uzd}{[ U_0^\dagger ]}

\title{Finite volume scaling of the electro-magnetic pion form factor in the $\epsilon$ regime}

\ShortTitle{Electro-magnetic pion form factor in the $\epsilon$ regime}

\author{Hidenori Fukaya\\
        Department of Physics, Osaka University, Toyonaka, Osaka 560-0043 Japan\\
        E-mail: \email{hfukaya@het.phys.sci.osaka-u.ac.jp}}

\author{\speaker{Takashi Suzuki}\\
       Department of Physics, Osaka University, Toyonaka, Osaka 560-0043 Japan\\
        E-mail: \email{suzuki@het.phys.sci.osaka-u.ac.jp}}

\abstract{
We consider finite volume effects on the electro-magnetic pion form factor near the
chiral limit, in the so-called $\epsilon$ regime. 
The pseudoscalar-vector-pseudoscalar three-point function is calculated in the $\epsilon$ expansion 
of chiral perturbation theory to the next-to-leading order. 
In the $\epsilon$ regime, finite volume effects are non-perturbatively large in general. 
However, we find a way to remove its dominant part, by inserting momenta to the correlators, 
and taking an appropriate ratio of them. The subleading contribution is, then, 
shown to be perturbatively small, and one can extract the form factor 
as in a similar way to that in the $p$ regime.
}

\FullConference{31st International Symposium on Lattice Field Theory - LATTICE 2013\\
		July 29 - August 3, 2013\\
		Mainz, Germany}

\begin{document}

\section{Introduction}

The pion form factor is one of the
fundamental low-energy quantities in QCD.
In terms of chiral perturbation theory (ChPT),
it is related to the low-energy constants (LECs) 
at the next-to-leading order \cite{Gasser1, Gasser2}.
However, it is still a challenge for lattice QCD to 
fully understand the low-energy behavior of the pion form factors.
In fact, all the lattice data so far simulated 
show lower values of the pion charge radius, 
which is obtained from the electro-magnetic form factor of the pions,
than that of the experiment (see a recent summary in \cite{Brandt:2013ffb}).

The lower values of the pion charge radius in lattice QCD 
are not surprising, since ChPT predicts a logarithmic divergence 
towards the pion mass zero limit,
and we can assume that the enhancement has just not appeared yet, 
with our simulated pion masses.
But in order to confirm such a logarithmic curve
from the first principle, 
a lattice QCD simulation near the physical point of the pion mass is essential.

Simulating QCD near the physical pion mass requires a 
large numerical cost to make the finite volume effects under control.
It is often stated that the pion mass multiplied by the lattice size,
$m_{\pi} L$ should be greater than 4 \cite{Aoki:2013ldr}.
If we want to employ the overlap fermions
or domain-wall fermions to keep a good chiral symmetry,
which may be another essential point for reproducing 
the chiral logarithm,
this requirement is difficult with currently available computational resources.

In this work, we would like to propose an alternative direction:
to find lattice observables which have small sensitivity to the volume size.
The key issue is how to reduce the contribution from the pion zero mode.
In fact, the above criteria $m_{\pi} L=4$, comes from 
a naive estimate for the zero-momentum pion's propagation
wrapping around the lattice, $\exp(-m_\pi L)$, as $m_\pi$ is its energy.
If we can remove the effect of the zero-mode, 
and replace the pion mass by some higher energy 
in the above estimate: $E_\pi L$, our lattice size $L$ can be taken smaller.

To this end, we consider the ``worst'' case:
the so-called $\epsilon$ regime \cite{Gasser:1986vb} where $m_\pi L<1$.
In the $\epsilon$ regime, 
the finite volume effects are $\sim 100$ \% in general,
but it is possible to quantify them within the $\epsilon$ expansion of ChPT.
In this work, we calculate the pseudoscalar-vector-pseudoscalar three-point function
in the $\epsilon$ expansion, and find a way to cancel 
the zero-mode's effect by inserting the momentum (or taking subtraction),
and making appropriate ratios of them with different momenta. 
We show that the remaining finite volume effect from non-zero modes
is perturbatively small $\sim {\cal O}(1/F^2\sqrt{V})$,
where $F$ denotes the pion decay constant.
Unlike many other examples in the $\epsilon$ regime,
our method does not use any peculiarity of the $\epsilon$ expansion,
and makes the analysis almost the same as in the $p$ expansion.
Since this method works in the ``worst'' case, 
the application to the $p$ regime should be straightforward.

\section{The $\epsilon$ expansion of ChPT}

We consider ChPT in an Euclidean finite volume 
$V=TL^3$ with the periodic boundary condition in every direction.
The chiral Lagrangian \cite{Gasser1,Gasser2} is given by
\begin{eqnarray}
\label{eq:ChL}
\mathcal{L}_{\rm ChPT}
=
\frac{F^2}{4} {\rm Tr} \left[ \left( \partial_\mu U(x) \right)^\dagger \left( \partial_\mu U(x) \right) \right]
-
\frac{\Sigma}{2} {\rm Tr} \left[ \mathcal{M}^\dagger U(x) + U^\dagger (x) \mathcal{M} \right]
+
\cdots,
\end{eqnarray}
where $U(x)$ is the chiral field taken as an element of the group $SU(N_f)$. 
$\Sigma$ denotes the chiral condensate and $F$ is the pion decay constant both in the chiral limit. 
Here, the higher order terms are not shown but exist, which is expressed by
the ellipses. For simplicity, we consider a diagonal
quark mass matrix $\mathcal{M} = {\rm diag} (m,m,m,\cdots)$.

\par

In the $\epsilon$ regime ($m_\pi V^{1/4} \ll 1$) \cite{Gasser:1986vb}, 
we need to integrate 
the zero-momentum mode of pions exactly, since its fluctuation becomes 
non-perturbatively large.
Thus, separating the zero-mode $U_0\in SU(N_f)$ from the others,
we parametrize the chiral field as
\begin{eqnarray}
U(x)
=
U_0 \, \ex \left( \frac{i \sqrt2 }{F} \xi(x) \right),
\end{eqnarray}
where $\xi(x) = T^a \xi^a(x)$ denotes the non-zero momentum modes.
Here, $T^a$'s are the generators of $SU(N_f)$ group, of which normalization 
is determined by ${\rm Tr} [ T^a T^b ] = \frac{1}{2} \delta^{ab}$.
Since the constant mode is denoted by $U_0$, a constraint on $\xi(x)$
\begin{eqnarray}
\label{eq:constraint}
\int d^4x \, \xi (x) = 0,
\end{eqnarray}
should be always satisfied to avoid the double-counting of the zero-mode.

Now let us expand ChPT according to the counting rule
\begin{eqnarray}
U_0 &\sim& \mathcal{O}(1),\;\;\;\;\;
\epsilon \sim \partial_\mu \sim \frac{1}{V^{1/4}} \sim m_\pi^{1/2} \sim m^{1/4} \sim \xi(x).
\end{eqnarray}
The Chiral lagrangian Eq.~(\ref{eq:ChL}) in this $\epsilon$ expansion is given by
\begin{eqnarray}
\mathcal{L}_{\rm ChPT}
&=&
-\frac{\Sigma}{2} {\rm Tr} \left[ \mathcal{M}^\dagger U_0 + U_0^\dagger \mathcal{M} \right]
+
\frac{1}{2} {\rm Tr} \left[ \partial_\mu \xi \partial_\mu \xi \right] (x)
+
\frac{\Sigma}{2F^2} {\rm Tr} \left[ \left( \mathcal{M}^\dagger U_0 + U_0^\dagger \mathcal{M} \right) \xi^2 \right] (x)
+
\cdots.\nonumber\\
\end{eqnarray}
Namely, we have to consider a hybrid system of bosonic field $\xi$ and a
matrix theory $U_0$, which weakly interact.

For $\xi(x)$ fields, it is not difficult to perform the Gaussian integrals
in terms of the ``propagator'',
\begin{eqnarray}
\langle \xi_{ij}(x) \xi_{kl} (y) \rangle_\xi
&=&
\delta_{il} \delta_{jk} \bar{\Delta}(x-y) - \delta_{ij} \delta_{kl} \frac{1}{N_f} \bar{\Delta}(x-y),
\end{eqnarray}
where the second term comes from the constraint ${\rm Tr}\xi =0$.
It is important to note that
\begin{eqnarray}
\bar{\Delta} (x)
&\equiv&
\frac{1}{V} \sum_{p \neq 0} \frac{e^{ipx}}{p^2},
\end{eqnarray}
denotes the propagation of the massless bosons, except that
the zero-mode contribution is absent 
in the sum over momentum
$p = 2\pi (n_t/T,\, n_x/L,\, n_y/L,\, n_z/L),$
with integers $n_\mu$.

On the other hand, the integral over the zero-modes (denoted by $\langle \cdots \rangle_{U_0}$) 
has to be non-perturbatively treated as a matrix integral,
which often requires non-trivial mathematical techniques. 
The zero-mode integration is in general expressed by the Bessel functions \cite{Splittorff-Verbaarschot},
which makes the correlator look quite different from that in the conventional $p$ regime.

This special feature of the zero-mode contribution in the $\epsilon$ regime 
has been used as an advantage to extract the physical quantities 
which are sensitive to the finite volume.
For example, the chiral condensate $\Sigma$ can be cleanly determined \cite{Fukaya:2009fh}
since the other higher-order LEC's give smaller contribution 
in the $\epsilon$ regime than that in the $p$ regime.
In this work, however, we show another direction : how to reduce such a
peculiarity of the $\epsilon$ regime.
Since the zero-mode contribution always appears as a position independent constant,
we can remove it with simple manipulations.


\section{Removing the dominant finite volume effects in the $\epsilon$ expansion}

As we have stressed in the previous sections, 
the dominant finite volume effect on the correlators
$\langle O(x_1)O(x_2)O(x_3)\cdots \rangle$ comes from the zero-mode of the pion. 
Since the zero-mode itself has no $x_i$ dependence, 
its effects always appear as $x_i$ independent constant terms,
or overall constant factors on $x_i$-dependent part, which come from
the interaction with the non-zero modes.
Therefore, it is always possible to remove them from the LO contribution
of the $\epsilon$ expansion.
To illustrate this, let us consider the simplest example:
the two point function.

The two point correlation function of the pseudoscalar operators can be expressed by
\begin{eqnarray}
\label{eq:twopt}
\langle P(x) P(y) \rangle
&=&
\mathcal{A}
+
\mathcal{B}
\frac{1}{V} \sum_{q\neq 0} \frac{e^{iq(x-y)}}{q^2}
+
\cdots,
\end{eqnarray}
where $\mathcal{A}$ and $\mathcal{B}$ denote the zero-mode's contribution, 
and ellipses imply higher order terms. 

First, let us remove $x$ or $y$-independent part $\mathcal{A}$. 
It is not difficult to see that any constant contribution cannot survive 
in the Fourier transform,
$f(x_0 ; {\bf p})
\equiv
\int d^3 x \, e^{-i {\bf p} \cdot {\bf x}} f(x),$
where ${\bf p \neq 0}$ is the three dimensional momentum. 
Namely, we have 
\begin{eqnarray}
\label{eq:inserted}
\langle P(x_0;{\bf p}) P(y_0;{\bf p'}) \rangle
&=&
\mathcal{B}
\frac{L^3}{T}
\delta_{{\bf p},{\bf -p'}}
\sum_{q^0} \frac{e^{iq^0(x_0-y_0)}}{(q^0)^2 + {\bf p}^2}
+
\cdots,
\end{eqnarray}
where $x_0$ or $y_0$ denote the temporal element of $x$ and $y$, respectively.
It is physically reasonable that the correlator having a momentum 
is less sensitive to the finite volume effects.

Even with ${\bf p = 0}$, it is possible to remove $\mathcal{A}$ by a subtraction of 
the operators at different time-slices:
$\Delta f ( {x_0 ; {\bf 0}} )
\equiv
f(x_0; {\bf 0}) - f(x_0^{\rm ref}; {\bf 0}),$
where $x_0^{\rm ref}$ is the arbitrary reference time,
provided that 
$x^{\rm ref}_0$ is satisfactory larger than the inverse QCD scale $1/\Lambda_{\rm QCD}$,
to avoid a contamination from the excited states.
More explicitly, we have 
\begin{eqnarray}
\label{eq:subtracted}
\langle \Delta P(x_0;{\bf 0}) P(y_0;{\bf 0}) \rangle
&=&
\mathcal{B}
\frac{L^3}{T}
\sum_{q^0 \neq 0}
\frac{e^{iq^0(x_0 - y_0)}}{(q^0)^2}
+
\cdots.
\end{eqnarray}

Next, let us remove the over-all constant $\mathcal{B}$ of the $x-y$ dependent part.
Noting that the same $\mathcal{B}$ is shared among the correlators with different momenta,
one can easily remove this by taking a ratio of them, for example,
\begin{eqnarray}
\label{eq:ratio}
\frac{\langle P(x_0;{\bf p}) P(y_0;{\bf p'}) \rangle}
{\langle \Delta P(x_0;{\bf 0}) P(y_0;{\bf 0}) \rangle}
&=&
\delta_{{\bf p},{\bf -p'}}
\frac{\sum_{q^0} \frac{e^{iq^0(x_0-y_0)}}{(q^0)^2 + {\bf p}^2}}
{\sum_{q^0 \neq 0}
\frac{e^{iq^0(x_0 - y_0)}}{(q^0)^2}}
+
\cdots.
\end{eqnarray}
Note that the higher order terms (denoted by elliplses) 
still have the zero-mode's effects, but they should be suppressed 
typically by $1/4\pi F^2V^{1/2}$. 

We note that this strategy is always applicable to any correlation functions.
In the above simplest example, there remains no important low-energy QCD
quantity unfortunately.
However, this is not the case in our main target, the three-point functions in the next section.

\section{Pseudoscalar-Vector-Pseudoscalar three-point function}
Now let us consider the pseudoscalar-vector(in the temporal direction)-pseudoscalar three-point function,
which is a relevant correlator to determine the pion charge radius in lattie QCD.
We calculate it to the next-to-leading order (or 1-loop in $\xi$ fields),
and find that the following three expressions are useful:
\begin{eqnarray}
\label{eq:pvp}
\langle P(x_0:{\bf -p}_f) V_0(y_0:{\bf q}) P(z_0:{\bf p}_i) \rangle
&=&
\frac{L^3 \Sigma_{\rm eff}^2}{4F_{\rm eff}}
\langle \mathcal{C} (U_0) \rangle_{U_0}
\delta_{{\bf q}, {\bf p}_f - {\bf p}_i}
F_V(q_0,{\bf q})
\nonumber \\ && \hspace{-40pt}
\times
\left[ iE({\bf p}_i) c({\bf p}_f, t) s({\bf p}_i, t') + iE({\bf p}_f) s({\bf p}_f, t) c({\bf p}_i, t') \right]
+
\cdots,\\
\label{eq:dpvp}
\langle \Delta P(x_0:{\bf 0}) V_0 (y_0:{\bf q}) P(z_0:{\bf p}_i) \rangle
&=&
\frac{L^3 \Sigma_{\rm eff}^2}{4F_{\rm eff}}
\langle \mathcal{C} (U_0) \rangle_{U_0}
\delta_{{\bf q}, - {\bf p}_i}
F_V(q_0,{\bf q})
\nonumber \\ && \hspace{-38pt}
\times
\left[ iE({\bf p}_i) \Delta c({\bf 0}, t) s({\bf p}_i, t') + iE({\bf 0}) \Delta s({\bf 0}, t) c({\bf p}_i, t') \right]
+
\cdots,\\
\label{eq:dpvdp}
\langle \Delta P(x_0:{\bf 0}) V_0 (y_0:{\bf q}) \Delta P(z_0:{\bf 0}) \rangle
&=&
\frac{L^3 \Sigma_{\rm eff}^2}{4F_{\rm eff}}
\langle \mathcal{C} (U_0) \rangle_{U_0}
\delta_{{\bf q},  {\bf 0}}
F_V(q_0,{\bf q})
\nonumber \\ && \hspace{-38pt}
\times
\left[ iE({\bf 0}) \Delta c({\bf 0}, t) \Delta s({\bf 0}, t') + iE({\bf 0}) \Delta s({\bf 0}, t) \Delta c({\bf 0}, t') \right]
+
\cdots.
\end{eqnarray}

Here, $F_V(q_0,{\bf q})$ is our target quantity, the vector form factor of the pion,
including some (perturbative) finite volume effects.
But first, we discuss the other parts to be removed.
Note that the low-energy constants include the one-loop corrections:
$\Sigma_{\rm eff}=\Sigma\left(1+\frac{(N_f^2-1)\beta_1}{N_f F^2V^{1/2}}\right)$, 
$F_{\rm eff}=F\left(1+\frac{N_f\beta_1}{2F^2V^{1/2}}\right)$ with 
the shape-coefficient $\beta_1$ (See Ref.~\cite{Hasenfratz:1989pk}). 
The dependence on $t = x_0 - y_0$ (and $t' = y_0 - z_0$) is expressed by
\begin{eqnarray}
c({\bf p},t)
=
\frac{\cosh \left[ E({\bf p}) (t-T/2)) \right]}
{2E({\bf p}) \sinh \left[ E({\bf p}) (t-T/2)) \right]},
\ \ \ 
s({\bf p},t)
=
\frac{\sinh \left[ E({\bf p}) (t-T/2)) \right]}
{2E({\bf p}) \sinh \left[ E({\bf p}) (t-T/2)) \right]}.
\end{eqnarray}
The zero-mode's contribution is contained in\footnote{
We use the same notation as the one in Ref.~\cite{Aoki-Fukaya}.
} 
\begin{eqnarray}
\label{eq:cuzero}
\mathcal{C}(U_0)
&=&
2 + 2\uz_{11}\uz_{22}  + 2\uzd_{11} \uzd_{22} + \uz_{11} \uzd_{11} + \uz_{22} \uzd_{22},
\end{eqnarray}
and the ``energy'',
\begin{equation}
E({\bf p}) = \sqrt{{\bf p}^2+\langle M_{\epsilon}^2(U_0)\rangle_{U_0}},
\end{equation}
where $M_{\epsilon}^2(U_0)$ is the ``pion mass'' in the $\epsilon$ expansion.
We don't give here the explicit form of $M_{\epsilon}^2(U_0)$ because
it is lengthy, and irrelevant in extracting $F_V(q_0,{\bf q})$,
as shown below.
It is only important to note that $M_{\epsilon}^2(U_0)$ converges to
the conventional $M_\pi^2$ in the $V\to \infty$ limit.

Next, let us remove the zero-mode contribution $\langle \mathcal{C}(U_0) \rangle_{U_0}$, by
making two ratios,
\begin{eqnarray}
\label{eq:ratio1}
R^1(t,t';{\bf p}_f,{\bf p}_i)
&&\equiv
\frac{\langle P (x_0 : -{\bf p}_f) V_0(y_0 : {\bf p}_f - {\bf p}_i) P(z_0 : {\bf p}_i) \rangle }
{\langle \Delta P (x_0 : {\bf 0}) V_0(y_0 : {\bf 0}) \Delta P(z_0 : {\bf 0}) \rangle}
\nonumber \\
&&\hspace{-0.5in}=
F_V(q_0, {\bf q})
\frac{E({\bf p}_i) c({\bf p}_f , t) s({\bf p}_i , t') + E({\bf p}_f) s({\bf p}_f , t) c({\bf p}_i , t') }
{E({\bf 0}) \Delta c({\bf 0} , t) \Delta s({\bf 0} , t') + E({\bf 0}) \Delta s({\bf 0} , t) \Delta c({\bf 0} , t')}
+\mathcal{O}(e^{-ET/2}),
\\
\label{eq:ratio2}
R^2(t,t';{\bf p}_f={\bf 0},{\bf p}_i)
&&\equiv
\frac{\langle \Delta P (x_0 : {\bf 0}) V_0(y_0 : {\bf p}_f - {\bf p}_i) P(z_0 : {\bf p}_i) \rangle }
{\langle \Delta P (x_0 : {\bf 0}) V_0(y_0 : {\bf 0}) \Delta P(z_0 : {\bf 0}) \rangle}
\nonumber \\
&&\hspace{-0.5in}=
F_V(q_0, {\bf q})
\frac{E({\bf p}_i) c({\bf p}_f , t) s({\bf p}_i , t') + E({\bf p}_f) s({\bf p}_f , t) c({\bf p}_i , t') }
{E({\bf 0}) \Delta c({\bf 0} , t) \Delta s({\bf 0} , t') + E({\bf 0}) \Delta s({\bf 0} , t) \Delta c({\bf 0} , t')}
+\mathcal{O}(e^{-ET/2}).
\end{eqnarray}

In these ratios, the zero-mode's contribution as well as the other peculiar expressions in the $\epsilon$
such as $\Sigma_{\rm eff}$, and $F_{\rm eff}$ are all cancelled except for those in 
$E({\bf p})$ through $\langle M_{\epsilon}^2(U_0)\rangle_{U_0}$.
However, as $t$ and $t^\prime$ dependences are explicitly known, 
we can fit the lattice data to the above ratios, 
treating $\langle M_{\epsilon}^2(U_0)\rangle_{U_0}$ as a free parameter, to extract $F_V(q_0, {\bf q})$.
Therefore, we don't need the explicit form of $\langle M_{\epsilon}^2(U_0)\rangle_{U_0}$.
We also find that the NLO contributions which are not proportional to $F_V(q_0, {\bf q})$
are exponentially small (as denoted by $\mathcal{O}(e^{-ET/2})$)\footnote{
When we strictly apply the $\epsilon$ expansion of ChPT, those $\mathcal{O}(e^{-ET/2})$ terms cannot be
neglected until we numerically confirm that these contribution is exponentially small and negligible. 
}.
Then the remaining finite volume effects (from the non-zero modes)
are only those within $F_V(q_0, {\bf q})$, which should be perturbatively small.

\section{Remaining finite volume effects}

Finally, let us consider the remaining perturbative finite volume effects in $F_V(q_0, {\bf q})$.
Although all the contributions from the zero-mode $U_0$ have been already removed, 
it still needs a lengthy expression.
Here let us just express it\footnote{
The explicit form of $\Delta F_V (q_0,{\bf q})$ will be shown in our paper \cite{SuzukiFukaya}.
} by
\begin{eqnarray}
\Delta F_V (q_0,{\bf q})
&\equiv&
F_V (q_0,{\bf q}) - F_V^{\infty} (q^2),
\end{eqnarray}
where $F_V^{\infty} (q^2)$ is the well-known result for the pion form factor within ChPT
(in the chiral limit)
\begin{eqnarray}
F_V^\infty(q^2) &=& 
1 - \frac{2 L^r_9(\mu_{sub})}{F_{\rm eff}^2}q^2-\frac{N_f}{2F_{\rm eff}^2}
\frac{1}{16\pi^2}
\left[
-\frac{1}{6}q^2 \ln \frac{q^2}{\mu_{sub}^2}+\frac{5}{18}q^2
\right],
\end{eqnarray}
where $L^r_9(\mu_{sub})$ is the renormalized LEC at the scale $\mu_{sub}$.

\begin{figure}[t]
\begin{center}
\includegraphics[bb=0 0 968 653,width=9.5cm]{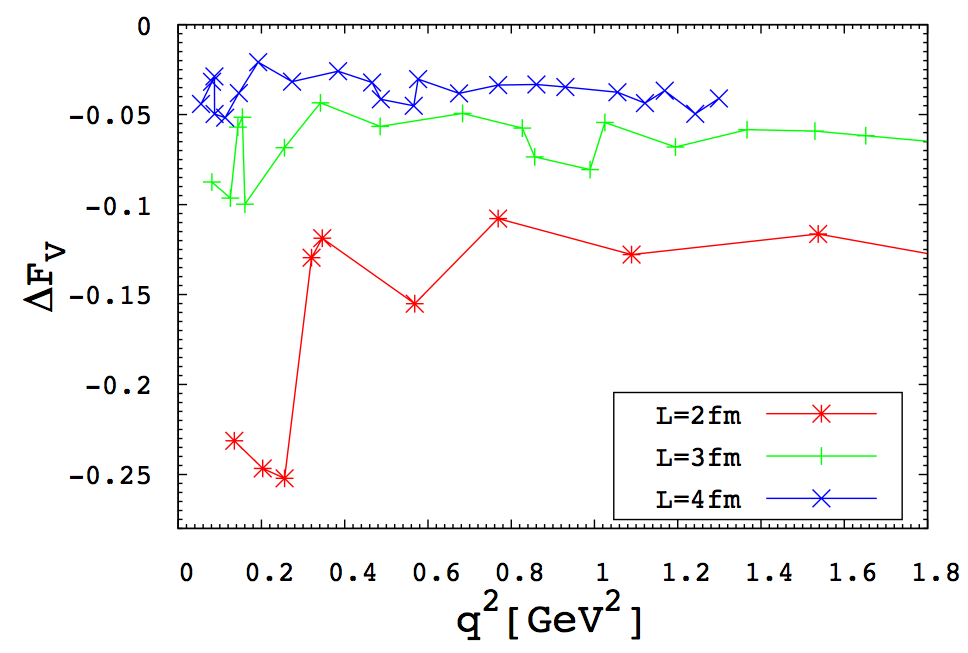}
\caption{Numerical estimates for $\Delta F_V$.}
\label{fig:DeltaF}
\end{center}
\end{figure}

Figure~\ref{fig:DeltaF} shows our numerical estimates for $\Delta F_V (q_0,{\bf q})$ 
as a function of ${\bf q}={\bf p}_f - {\bf p}_i$, and 
$q_0=i\left(\sqrt{{\bf p}_f^2+M_\pi^2}-\sqrt{{\bf p}_i^2+M_\pi^2}\right)$ where we choose $M_\pi=135$MeV,
and $F_{\rm eff}=92.2$ MeV. 
We take three values of $L= 2,3,4 \, [{\rm fm}]$ with $T=2L$, and $N_f = 2$. 
As $F^{\infty} (q^2)$ is $\mathcal{O}(1)$ quantity, one can see that 
the finite volume effect is perturbatively small.
In particular, it is important to note that it is less 10\% already at $L=3$ fm.

As a final remark of this section, let us comment on the contribution from the heavier particles.
It is known in lattice QCD simulations that ChPT estimate for  $F_V^\infty(q^2)$
is not enough but one needs to include the effect from the rho-meson resonance.
However, even in this case, as the rho meson or heavier hadrons have 
negligibly smaller sensitivity to the finite volume
than that of the pion, $\Delta F_V (q_0,{\bf q})$ should still be a 
good estimate for the finite volume effect on the pion form factor.

\section{Summary}

We have considered finite volume effects on 
the electro-magnetic pion form factor near the chiral limit.
The pseudoscalar-vector-pseudoscalar three-point function,
which is a relevant quantity in lattice QCD simulations to extract the form factor, 
was calculated in the $\epsilon$ expansion 
of chiral perturbation theory to the next-to-leading order. 

In the $\epsilon$ regime, as the correlation length of the pion 
exceeds the volume size,
its finite volume effects become ${\cal O}(1)$ in general. 
However, we have shown that it is possible to remove its dominant part, 
by inserting momenta to the correlators, or making subtraction 
with the one at a different time slice,
and taking an appropriate ratio of them. 

The subleading contribution from the non-zero momentum modes of the pion 
was, then,  shown to be perturbatively small.
This method allows us to extract the pion form factor 
as in a similar way to that in the $p$ regime.
Especially, we would like to stress that we don't need any 
Bessel functions in the analysis.
In fact, our result (at tree-level) was already tried on a small lattice,
and a large value of the pion charge radius was reported \cite{Fukaya:2012dla}.
Since this method works in the ``worst'' case in the $\epsilon$ regime,
and does not need any special feature of the $\epsilon$ expansion,
the application to the $p$ regime should be straightforward.

We thank S.~Aoki, S.~Hashimoto, and T.~Onogi for useful discussions.
The work of HF is supported in part by the Grant-in-Aid of the Japanese Ministry of Education (No. 25800147).

\end{document}